\documentstyle{elsart}
\input psfig

\begin{document}
\begin{frontmatter}
\title{Vortex Lattice Depinning vs. Vortex Lattice Melting: a pinning-based explanation of the equilibrium 
magnetization jump}

\author[leipzig]{Y. Kopelevich\thanksref{kope}} and
\author[leipzig]{P. Esquinazi}
\address[leipzig]{Department of Superconductivity and Magnetism, 
Institut  f\"ur Experimentelle Physik II, 
Universit\"at Leipzig, Linn\'estr. 5, D-04103 
Leipzig, Germany}
\thanks[kope]{On leave from Instituto de Fisica, Unicamp, 
13083-970 Campinas, Sao Paulo, Brasil. Supported by the 
Deutsche 
Forschungsgemeinschaft under DFG IK 24/B1-1,
Project H, and by CAPES proc. No. 077/99.}

\begin{abstract}
In this communication we argue that the Vortex Lattice Melting scenario fails to explain several 
key experimental
results published 
in the literature. From a careful analysis of these  results 
we conclude that the Flux Line Lattice (FLL) does not melt along a material- and sample-dependent 
boundary $H_j(T)$ but the opposite, it de-couples from the superconducting matrix becoming 
more ordered. When the FLL depinning is sharp, the difference between the 
equilibrium magnetization $M_{eq}(T,H)$ of the pinned and unpinned FLL leads to the observed
step-like change $\Delta M_{eq}(T,H)$. We demonstrate that the experimentally obtained 
$\Delta M_{eq}(T,H)$ can be well accounted for by a variation of the pinning efficiency 
of vortices along the $H_j(T)$ boundary.
\end{abstract}
\end{frontmatter}


The phenomenological description of superconductors is based on the 
knowledge of their magnetic field-temperature $(H-T)$ phase diagram. The equilibrium behavior 
of conventional type-II superconductors in an applied magnetic field is well known: At fields 
below the lower critical field $H_{c1}(T)$, the superconductor is in the Meissner-Ochsenfeld phase in 
which surface currents screen the magnetic field from the interior of the sample. Above $H_{c1}(T)$ the 
field penetrates the superconductor in the form of a lattice of vortices, the so-called 
Abrikosov vortex lattice. This flux-line-lattice (FLL) persists up to the upper critical field $H_{c2}(T)$ 
where 
superconductivity vanishes in the bulk of the sample. 

In high-temperature superconductors (HTS), however, due to strong 
thermal fluctuations, a first-order phase transition 
of the FLL to a liquid-like state, the ``melting'' of the FLL, has been predicted to occur \cite{nelson,brandt}
 well below $H_{c2}(T)$. Since then, an enormous amount of experimental and theoretical work 
has been done trying to find this, or other more sophisticated transitions and to extend the original 
theoretical treatment\cite{blatter,brandt2}. At the beginning of these research activities it was claimed 
that the melting transition of the FLL in HTS manifests itself at the damping peak of vibrating HTS in 
a magnetic field\cite{gammel}. However, this interpretation was controversial \cite{brandt3}.
We know nowadays that the damping peaks in vibrating superconductors 
attributed to the melting  transition \cite{gammel} can be explained quantitatively
assuming thermally activated depinning and the diffusive motion of the FLL under a 
small perturbation generated by the vibration of the sample \cite{ziese}. To find the true 
experimental evidence for the melting transition of the FLL is by no means simple: Because 
the FLL interacts with the superconducting matrix through pinning centers (atomic lattice defects, surface 
barriers, etc.) every property of the FLL one measures will be influenced by the pinning and, therefore, 
no direct and straightforward proof of the melting phase transition can be achieved. 

Several years after the above cited first experimental attempt, a striking jump in the equilibrium magnetization 
in Bi$_2$Sr$_2$CaCu$_2$O$_8$ (Bi2212) single crystals has been measured using a SQUID\cite{pastoriza} 
as well as sensitive micro-Hall sensors\cite{zeldov}. This magnetization jump, which was interpreted as 
a first-order transition of the FLL, lies at more than one order of magnitude lower fields than the 
thermally activated depinning line measured with vibrating superconductors\cite{gammel,ziese}. 
These interesting results\cite{pastoriza,zeldov}are important because if the melting transition would 
be of the first order, a discontinuous change in the equilibrium magnetization $M_{eq}(T,H)$ at the 
transition is expected. The step-like increase of $M_{eq}(T,H)$ (a decrease in absolute value) with 
increasing magnetic field and temperature has been also observed in YBa$_2$Cu$_3$O$_7$ (Y123) 
single crystals\cite{welp}.
 
Apart from the clearly defined magnetization jump, another important fact was revealed by 
the experiments. It has been found that the line $H_j(T)$ along which $M_{eq}(T,H)$ jumps, 
and the temperature dependence of the field $H_{\rm SMP}(T)$ where an anomalous 
maximum in the width of the magnetization hysteresis loop takes place (the so-called ``second 
magnetization peak'' 
(SMP)), define a unique boundary on the $H-T$ plane for a given sample,\cite{khaykovich,nishi}
 demonstrating their intimate relationship. In order to account for this behavior, a second-order phase 
transition associated with the increase of the critical current density $j_c(T,H)$ was suggested as the 
origin of SMP based on a pinning-induced disordering of the FLL\cite{ertas,giller,vinokur}.
 
Recently, a similar SMP occurring at a temperature dependent field $H_{\rm SMP}(T) \ll H_{c2}(T)$ 
was also measured in conventional Nb superconducting films\cite{kope,esqui}. The studies of the 
SMP performed on Nb films,\cite{kope,esqui} Bi2212 single crystals,\cite{kope2} as well as on 
a non-cuprate isotropic single crystalline Ba$_{0.63}$K$_{0.37}$BiO$_3$ thick film\cite{galkin}
 provide a clear evidence that the SMP is not related to a critical current enhancement, but originates 
from a thermomagnetic instability effect and/or a non-uniform current distribution,\cite{gurevich} 
leading to the ``hollow'' in $M(H)$ at $H < H_{\rm SMP}(T)$. In agreement with the models, the 
SMP vanishes in all these superconductors when the lateral sample size becomes less 
than $\sim 100~\mu$m \cite{kope,esqui,kope2,galkin}. 
Because $100~\mu$m is much larger than all relevant vortex-pinning-related characteristic lengths, 
the strong influence of the sample geometry on the SMP cannot be explained by a change of 
the pinning efficiency of vortices. Besides, the results\cite{esqui,kope2} suggest that the interaction 
between vortices starts to dominate that between vortices and the matrix at $H > H_{\rm SMP}(T)$. 
We note that the results obtained in Bi2212 crystals\cite{kope2} are actually in good agreement 
with the second-order diffraction small-angle neutron scattering (SANS) experiments which revealed a 
well-ordered FLL at $H > H_{\rm SMP}(T)$ \cite{forgan}. Moreover, the formation of a more 
ordered FLL with increasing temperature due to thermal depinning has been found in the above mentioned 
SANS experiments\cite{forgan} at $H > H_{\rm SMP}(T)$ and for intermediate temperatures. 
A similar result was obtained by means of Lorentz microscopy in Bi2212 thin films near the low 
field - high temperature portion of the ``irreversibility line''\cite{harada}. 
The high resolution SANS measurements, recently reported for Y123 crystals\cite{johnson} also 
revealed a well defined FLL up to a field of 4 T and at low temperatures, i.e. above the 
$H_{\rm SMP}(T)$-line measured in similar crystals\cite{nishi}. 

Based on this experimental evidence and the intimate relationship between SMP- and the 
magnetization-jump-lines,\cite{khaykovich,nishi} we propose that the jump in $M_{eq}(T,H)$ 
results from a magnetic-field- and temperature-driven FLL depinning transition to a more {\em ordered}
 state of the FLL, effectively de-coupled from the atomic lattice. 

There exist already experimental\cite{puzniak,van,li} as well as theoretical\cite{bula} works 
that show that the interaction of vortices with pinning centers {\em increases} $|M_{eq}(T,H)|$, indicating 
clearly that pinning influences the thermodynamic, equilibrium properties of superconductors. If the 
FLL depinning is sharp,\cite{kos,wag} then one expects a step-like change of the equilibrium 
magnetization $|\Delta M_{eq}(T,H)| = |M_{eq}^{dis}(T, H) - M_{eq}(T, H)|$ along the 
$H_j(T)$ boundary. Here $|M_{eq}^{dis}(T, H)| > |M_{eq}(T, H)|$ is the absolute 
equilibrium magnetization in the presence of the quenched disorder which is measured below 
$H_j(T)$. We note further that the sharpness of the vortex depinning onset, irrespectively of 
the underlying mechanism, manifests itself as a sudden increase in the electrical resistivity at  $H_j(T)$, 
below which the vortex behavior is irreversible\cite{fuchs}.

In what follows we present a phenomenological approach to describe the magnetization jump observed 
at the depinning transition. The equilibrium magnetization of an ordered, unpinned FLL in the 
London regime and neglecting  fluctuations\cite{bula2} is given by the equation 
\begin{equation}
                  M_{eq} = - \frac{\phi_0}{2(4\pi\lambda)^2} \ln(\eta H_{c2}/H)\,,	                                  
\label{eq1}
\end{equation}
where $\lambda(T) \equiv \lambda_{ab}(T)$ is the in-plane London penetration depth, 
$\phi_0$ is the flux quantum, and $\eta$ is a parameter analogous to the 
Abrikosov ratio $\beta_A = <|\Psi|^4>/<|\Psi^2|>^2 (\Psi$ being the superconducting 
order parameter) that depends on the FLL structure\cite{abri}. 
We assume now that the $\Delta M_{eq}$ results from a change of the 
parameter $\eta$ due to a change in the vortex arrangement triggered by the 
interaction of the FLL with pinning centers. Hence, the magnetization jump can be written as 
\begin{equation}
                       \Delta M_{eq} = \frac{\phi_0}{2(4\pi \lambda)^2} \ln(\eta^{dis}/\eta_0)\,,
	\label{eq2}
\end{equation}
where the parameter $\eta^{dis} \ge \eta_0$ is related to the strength of the quenched 
disorder, and $\eta_0$ applies for the FLL above the $H_j(T)$. Therefore, it is reasonable 
to assume that $\eta^{dis}$ is proportional to the critical current density $j_c(T,H)$ which 
measures the vortex pinning strength. In fact, the correlation between $\Delta M_{eq}$ and 
the pinning of the FLL has been observed experimentally\cite{farell}. We emphasise that 
although the relationship between $\eta$ and $\beta_A$ is unknown, one expects an 
increase of the parameter $\eta$ in the vortex liquid state compared to that of 
the vortex solid, similar to the results obtained for $\beta_A$ \cite{hikami}.
\begin{figure}
\centerline{\psfig{file=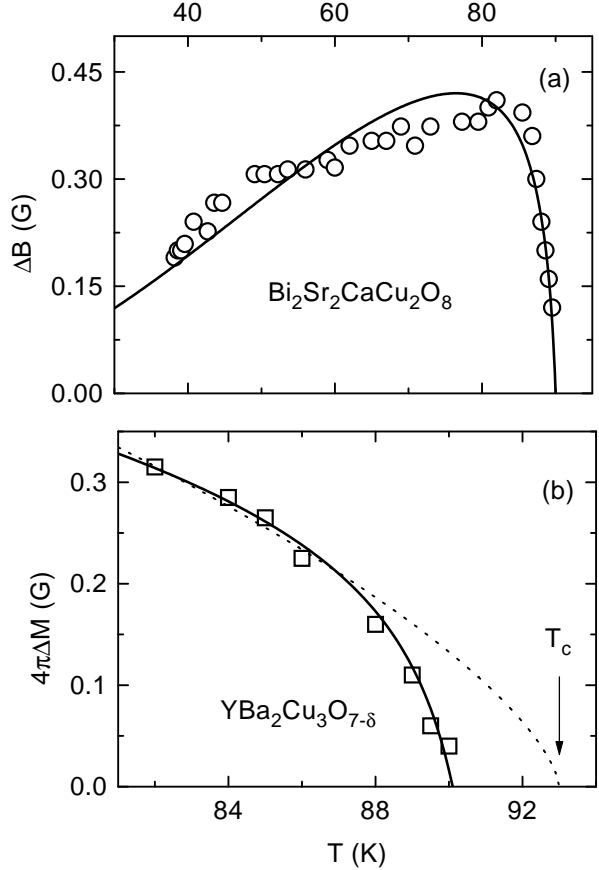,height=5.2in}}
\caption{The jump $\Delta B$ in the induction (a) and equilibrium magnetization 
$4\pi \Delta M$ (b) as a function of temperature measured for Bi$_2$Sr$_2$CaCu$_2$O$_8$ 
(Bi2212) \protect\cite{zeldov} and YBa$_2$Cu$_3$O$_{7-\delta}$ (Y123)\protect\cite{welp}
 single crystals, respectively. The solid lines are obtained from Eq.~(2) with 
$\eta^{dis} = \eta_0[1 + aj_{c0}(1 - T/T_0)^n/(H^{\alpha} + H_0)]$, with the fitting 
parameters $n = 1, \alpha = 1.8, H_0 = 200~$Oe$^{1.8}, T_0 = 90$~K, $aj_{c0} = 
9.3 \times 10^4~$Oe$^{1.8}$ for Bi2212 and $n = 1, \alpha = 1, H_0 = 3.5 \times 10^3~$Oe, 
$T_0 = 90.1~$K, $aj_{c0} = 2.6 \times 10^5$~Oe for 
Y123. Dotted line (b) corresponds to the equation $4 \pi \Delta M 
= 1.32 (1 - T/T_c)^{2/3} (Tc = 92.9~$K), according to the theoretical result from
melting theory \cite{dogson}. 
}
\end{figure}
  
We show in Fig.~1 the jumps of the induction $\Delta B(T,H)$ and of the magnetization 
$4\pi \Delta M(T, H)$ measured along the $H_j(T)$ boundaries in Bi2212 (a)\cite{zeldov}
 and Y123 (b)\cite{welp} single crystals, respectively. The difference in the 
behavior of $\Delta M_{eq}(T,H)$ in Bi2212 and Y123 crystals (see Fig.~1) can be easily 
understood noting that $H_j(T) < H^{\ast} = \phi_0/\lambda^2$ for Bi2212, 
whereas $H_j(T) \gg H^{\ast}$ in the case of Y123. At fields $H < H^{\ast}$ the FLL 
shear modulus exponentially decreases with field as $c_{66} \simeq 
(\epsilon_0/\lambda^2)(H\lambda^2/\phi_0)^{1/4}
\exp(-\phi_0/H\lambda^2)$, whereas at $H > H^{\ast}$ it is proportional to the 
field  $c_{66} \simeq (\epsilon_0/4\phi_0)H$ \cite{wag}.
 
The exponential decrease of $c_{66}$ with decreasing $H_j$ (increasing $T_j$) in the case 
of Bi2212, implies an enhancement of the interaction between vortices and the 
quenched disorder (pinning centers) which leads to an increase of $\Delta M_{eq}$ increasing $T_j$ 
(or decreasing $H_j$), see Fig.~1(a). As temperature tends to the critical one $T_c$, the 
pinning of vortices vanishes. Therefore, above a certain temperature, $\Delta M_{eq}(T,H)$ 
decreases with temperature and tends to zero. 

In the case of Y123, however, and due to a relatively weak field dependence of $c_{66}$ 
along the $H_j(T)$ line, the reduction of the vortex pinning efficiency with temperature 
is the dominant effect which explains the monotonous decrease of $\Delta M_{eq}(T,H)$ with 
temperature, see Fig.~1(b). We stress that the vanishing of the magnetization jump 
$(\Delta M_{eq}(T, H))$ at $T_0 \simeq 90~$K, i.e. approximately 3 K below the 
superconducting transition temperature $T_c = 92.9~$K\cite{welp} (similar result was obtained in 
another untwinned Y123 crystal\cite{schil}), can be also explained naturally by the 
effect of thermal fluctuations which smear out the pinning potential. On the other hand, the 
theory\cite{dogson} based on the FLL melting hypothesis predicts $\Delta
 M_{eq} \sim \phi_0/\lambda^2(T)$
 (dotted line in Fig.~1(b)) which implies the vanishing of $\Delta M_{eq}$ at $T_c$. While the 
FLL-melting theory\cite{dogson} requires different approaches in order to explain $\Delta M_{eq}(T, H)$ 
in Y123 and Bi2212, our analysis can equally well be applied to both weakly (Y123) and 
strongly (Bi2212) 
anisotropic superconductors.

In order to use Eq.~(2) to calculate the magnetization jump we need the relationship between $\eta^{dis}$  
and the critical current density $j_c(T, H)$ at the boundary $H_j(T)$, which is unknown at present. 
Nevertheless and as an illustration of our ideas we present here a simple fitting approach. The solid lines 
in Fig.~1(a,b) were obtained from Eq.~(2) with $\eta^{dis} = \eta_0[1 + aj_c(T, H)] = \eta_0[1 + 
aj_{c0}(1 - T/T_0)^n/(H^\alpha + H_0)]$ calculated at the boundary $H_j(T)$, where $T_0$ 
corresponds to the temperature at which $j_c = 0$, and $a, j_{c0}, H_0$ are 
model-dependent constants, $n$ and $\alpha$ are pinning related exponents. Note, that the 
here used $j_c(T,H)$ is a rather general expression which reflects the well-known experimental fact that 
the critical current density generally decreases with temperature and increasing field. 
In our fits (solid lines in Fig.~1) we have set the exponents $n = 1, \alpha = 1.8 (1)$, and 
used $\lambda(T) = 250 (1- T/T_c)^{-1/3}~$nm ($140 (1-T/T_c)^{-1/3}$~nm) for
 Bi2212 (Y123)\cite{kamal}; other fitting parameters close to those used give also satisfactory fits. 

Furthermore, within the here proposed physical picture we expect a reduction and ultimately the 
vanishing of $\Delta M_{eq}(T, H)$ if by applying external driving forces one de-couples the 
FLL from the matrix. Such effect was observed in Bi2212 crystals, indeed\cite{farell}. 
Also, within our picture we expect that the FLL remains in a more ordered, metastable state if the sample 
is field cooled as compared to the zero-field-cooled state. Several published results have indicated such 
a behavior, both directly (see, e.g. Ref.~\cite{oral}) and indirectly. Among them, the 
pioneer work\cite{pastoriza} which demonstrates that the magnetization jump at the irreversibility 
line can be much larger in the zero-field-cooled sample compared to the field-cooled one, providing 
a clear evidence that $\Delta M_{eq}(T, H)$ is essentially related to the competition between 
vortex-vortex and vortex-pinning centers interactions. Supporting the above ideas, the magnetic-field-driven 
transition from a disorder-dominated vortex state to a moving well-ordered FLL was observed in NbSe$_2$ 
low-$T_c$ layered superconductor \cite{pardo}.
 
Finally, we would like to point out that 
the equilibrium magnetization jump at a 
first-order depinning transition would imply the use of  the Clausius-Clapeyron relation. 
However, we are not aware of any prediction on the 
entropy change at the depinning transition which we could
use for comparison. Therefore,  we believe that it has little sense to comment here 
on the use of the Clausius-Clapeyron equation at this stage of our study. 

To summarise, based essentially on experimental facts we propose the magnetic-field and 
temperature-driven vortex-lattice-ordering transition as an alternative to the FLL melting
scenario in high-$T_c$ 
superconductors. Simple arguments allowed us to account for the equilibrium 
magnetization jump associated 
with the FLL depinning transition. This transition awaits for a rigorous theoretical treatment.

\ack
We acknowledge illuminating discussions with G. Blatter, E. H. Brandt, A.M. Campbell, G. Carneiro, 
F. de la Cruz, M. D\"aumling, R. Doyle, E.M. Forgan, D. Fuchs, A. Gurevich, B. Horovitz, 
M. Konczykowski, 
A. H. MacDonald, A. Schilling, T. Tamegai, V. Vinokur, E. Zeldov, and M. Ziese. This work 
was supported by 
the German-Israeli-Foundation for Scientific Research and Development and the Deutsche 
Forschungsgemeinschaft.


\begin{thebibliography}{9}
\bibitem{nelson} D. R. Nelson, Phys. Rev. Lett. {\bf 60}, 1973 (1988). 
\bibitem{brandt}E. H. Brandt, Phys. Rev. Lett. {\bf 63}, 1106 (1989). 
\bibitem{blatter}G. Blatter, M. V. Feigel'man, V. B. Geshkenbein, A. I. Larkin, and V. M. Vinokur, 
Rev. Mod. Phys. {\bf 66}, 1125 (1994).
\bibitem{brandt2}E. H. Brandt,  Rep. Prog. Phys. 58, 1465-1594 (1995). 
\bibitem{gammel}P. L. Gammel, L. F. Schneemeyer, J. V. Waszczak,  and D. Bishop,  
Phys. Rev. Lett. {\bf 61}, 1666-1669 (1988).
\bibitem{brandt3}E. H. Brandt, P. Esquinazi, and G. Weiss, Comment on Ref. 5. Phys. 
Rev. Lett {\bf 62}, 2330 (1989).
\bibitem{ziese}M. Ziese, P. Esquinazi, and H. F. Braun, Supercond. Sci. Technol. {\bf 7}, 
869 (1994).
\bibitem{pastoriza}H. Pastoriza, M. F.  Goffman,  A. Arribére and F. de la Cruz, Phys. Rev. 
Lett. {\bf 72}, 2951 (1994). 
\bibitem{zeldov}E. Zeldov, et al., Nature {\bf 375}, 373 (1995). 
\bibitem{welp}U. Welp et al., Phys. Rev. Lett. {\bf 76}, 4809 (1996). 
\bibitem{khaykovich}B. Khaykovich et al., Phys. Rev. Lett. {\bf 76}, 2555 (1996). 
\bibitem{nishi}T. Nishizaki, T.  Naito, and N. Kobayashi, Phys. Rev. B {\bf 58}, 11169 (1998).
\bibitem{ertas}D. Ertas, and D. R. Nelson,  Physica C {\bf 272}, 79, (1996).
\bibitem{giller}D. Giller et al., Phys. Rev. Lett. {\bf 79}, 2542 (1997). 
\bibitem{vinokur}V. Vinokur et al., Physica C {\bf 295}, 209 (1998). 
\bibitem{kope}Y. Kopelevich and P. Esquinazi, J. Low Temp. Phys. {\bf 113}, 1 (1998).
\bibitem{esqui}P. Esquinazi et al., Phys. Rev. B {\bf 60}, 12454 (1999). 
\bibitem{kope2}Y. Kopelevich, S. Moehlecke, J. H. S. Torres, R. Ricardo da Silva, and P. 
Esquinazi,  J. Low Temp. Phys. {\bf 116}, 261 (1999). 
\bibitem{galkin}A. Yu. Galkin et al., Solid State Commun. (2000, in press). 
We note that the occurrence of similar $H_{\rm SMP}(T)$-lines as well as the vanishing of the 
SMP for small enough samples in both layered and isotropic HTS indicate that the layer-decoupling 
scenario [see B. Horovitz,  Phys. Rev. B {\bf 60}, 9939 (1999)] does not apply. 
\bibitem{gurevich}A. Gurevich, V. M. Vinokur,   Phys. Rev. Lett. {\bf 83}, 3037 (1999). 
\bibitem{forgan}E. M. Forgan, M. T. Wylie, S. Lloyd, S. L. Lee,  and R. Cubitt, Proc. LT-21: 
Czechoslovak Journal of Physics {\bf 46}, 1571 (1996). 
\bibitem{harada}K. Harada et al.,  Phys. Rev. Lett. {\bf 71}, 3371 (1993). We note 
that the arguments used by these authors regarding 
the decrease in contrast of the signal for visualising the FLL by increasing field or
 temperature may, 
in principle, be applied to other techniques like neutron diffraction, muon spin 
rotation, and scanning Hall 
probe microscope (see Ref.~\protect\cite{oral}) studies.
\bibitem{johnson}S. T. Johnson  et al., Phys. Rev. Lett. {\bf 82}, 2792 (1999).
\bibitem{puzniak}R. Pu$\acute{\rm z}$niak, J. Ricketts, J. Sch\"utzmann, G. D. Gu, and N. Koshizuka, 
Phys. Rev. B {\bf 52}, 7042 (1995). 
\bibitem{van}C. J. van der Beek, M. Konczykowski, T. W. Li, P. H.  Kes, and W. 
Benoit,  Phys. Rev. B {\bf 54}, 792 (1996). 
\bibitem{li}Q.  Li et al., Phys. Rev. B {\bf 54}, 788 (1996). 
\bibitem{bula}L. N. Bulaevskii, V. M. Vinokur, and M. P. Maley, Phys. Rev. Lett. 
{\bf 77}, 936 (1996). 
\bibitem{kos}A. E. Koshelev,  and P. H. Kes, Phys. Rev. B {\bf 48}, 6539 (1993).
\bibitem{wag}O. S.  Wagner, G. Burkard, V. B. Geshkenbein, and G. Blatter, Phys. Rev. 
Lett. {\bf 81}, 906 (1998). 
\bibitem{fuchs}D. T. Fuchs et al., Phys. Rev. B {\bf 54}, 796 (1996). 
\bibitem{bula2}L. N. Bulaevskii, M.  Ledvij, and V. G. Kogan, Phys. Rev. Lett. {\bf 68}, 
3773 (1992). 
\bibitem{abri}A. A. Abrikosov, in {\it Fundamentals of the Theory of Metals}, North 
Holland (1998), pp. 414, 425. 
\bibitem{farell}D. E. Farell et al., Phys. Rev. B {\bf 53}, 11807 (1996). 
\bibitem{hikami}S. Hikami, A. Fujita,  and A. I. Larkin, Phys. Rev. B {\bf 44}, 10400 (1991). 
\bibitem{schil}A. Schilling et al., Nature {\bf 382}, 791 (1996).
\bibitem{dogson}M. J. W. Dodgson, V. B.  Geshkenbein, H. Nordborg, and G. Blatter,   
Phys. Rev. Lett. {\bf 80}, 837 (1998). 
\bibitem{kamal}S. Kamal et al., Phys. Rev. Lett. {\bf 73}, 1845 (1994). 
\bibitem{oral}A. Oral et al., Phys. Rev. Lett. {\bf 80}, 3610 (1998).
\bibitem{pardo}F. Pardo, F. de la Cruz, P. L. Gammel, E. Bucher, and D. J. Bishop, 
Nature {\bf 396}, 348 (1998).


\end{thebibliography}
\end{document}